\documentclass[a4paper,11pt]{article}
\usepackage{jcappub} 
\usepackage{lineno}

\usepackage{comment}
\usepackage{graphicx}
\usepackage{subfigure}
\usepackage{palatino}
\usepackage[commandnameprefix=always]{changes}
\usepackage{hyperref}
\hypersetup{colorlinks=true,linkcolor=blue,urlcolor=blue,citecolor=blue}
\usepackage[toc,page]{appendix}
\usepackage[normalem]{ulem}

\usepackage{orcidlink}
\usepackage{lipsum}
\usepackage{graphicx}
\usepackage{subfigure}
\usepackage{palatino}
\usepackage[table,xcdraw]{xcolor}
\usepackage{sans}
\usepackage{adjustbox}
\usepackage{latexsym}
\usepackage{amsmath}
\usepackage{amssymb}
\usepackage{amsfonts}
\usepackage{dcolumn}
\usepackage{bm}
\usepackage{tikz}
\usepackage{bigints}
\usepackage{array,tabularx,multirow,booktabs}
\usepackage[tracking=true]{microtype}
\SetTracking{}{500}
\SetTracking{encoding={*}, shape=sc}{40}
\UseRawInputEncoding 
\allowdisplaybreaks
\usepackage{adjustbox}
\usepackage{latexsym}
\usepackage{amsmath}
\usepackage{amssymb}
\usepackage{amsfonts}
\usepackage{dcolumn}
\usepackage{bm}
\usepackage{tikz}
\usepackage{bigints}
\usepackage{array,tabularx,multirow,booktabs}
\usepackage[tracking=true]{microtype}
\usepackage{color}

\def\1o2{{1\over2}}

\UseRawInputEncoding 
\allowdisplaybreaks

\title{\boldmath Null geodesics and shadows of slowly rotating wormholes immersed in dark matter halos}

\author[a,b]{Abdelghani Errehymy,\orcidlink{0000-0002-0253-3578}}
\author[c]{Abdullah Guvendi,\orcidlink{0000-0003-0564-9899}}
\author[d]{Semra Gurtas Dogan,\orcidlink{0000-0001-7345-3287}}
\author[e]{Omar Mustafa,\orcidlink{0000-0001-6664-3859}}

\affiliation[a]{Astrophysics Research Centre, School of Mathematics, Statistics and Computer Science, University of KwaZulu-Natal, Private Bag X54001, Durban 4000, South Africa}
\affiliation[b]{Center for Theoretical Physics, Khazar University, 41 Mehseti Str., Baku, AZ1096, Azerbaijan}
\affiliation[c]{Department of Basic Sciences, Erzurum Technical University, 25050, Erzurum, Türkiye}
\affiliation[d]{Department of Medical Imaging Techniques, Hakkari University, 30000, Hakkari, Türkiye}
\affiliation[e]{Department of Physics, Eastern Mediterranean University, 99628, G. Magusa, north Cyprus, Mersin 10 - Türkiye}

\emailAdd{abdelghani.errehymy@gmail.com (Corr. Auth.)}
\emailAdd{abdullah.guvendi@erzurum.edu.tr}
\emailAdd{semragurtasdogan@hakkari.edu.tr}
\emailAdd{omar.mustafa@emu.edu.tr}

\abstract{We study slowly rotating traversable wormholes embedded in realistic galactic dark matter halos, including Navarro-Frenk-White (NFW), Bose-Einstein Condensate (Thomas-Fermi, TF), and pseudo-isothermal (PI) profiles. Using the Teo-type rotating wormhole metric, we construct shape functions from halo density distributions and analyze the resulting geometrical properties, such as throat structure, flaring-out conditions, and violations of the null energy condition. We examine null geodesics, effective potentials, photon spheres, and Lense-Thirring (LT) precession, highlighting differences between cuspy and cored halo models. Finally, we calculate wormhole shadows, observing that cuspy NFW halos tend to produce smaller, asymmetric shadows, while cored TF and PI halos yield smoother, nearly circular silhouettes. The findings provide a theoretical characterization of photon dynamics and shadow morphology in wormholes embedded within different dark matter environments. }

\keywords{Rotating wormhole spacetimes; Null geodesics; Photon orbits and stability; Gravitational lensing; Ergoregions and optical horizons}

\begin{document}
\maketitle
\flushbottom

\section{Introduction}

\setlength{\parindent}{0pt}

The idea of a wormhole goes back to Flamm in 1916 \cite{flamm1916}, and later, in 1935, Einstein and Rosen proposed the existence of traversable wormholes, now called Einstein-Rosen bridges \cite{einstein1935}. The term \textit{wormhole} was introduced by Wheeler while discussing \textit{geons} \cite{Misner:1957mt}. Wormholes are thought of as shortcuts through spacetime, connecting two different points. So far, no one has been able to prove their existence experimentally; they remain purely mathematical concepts. Wheeler later demonstrated that wormholes are inherently unstable, preventing even photons from passing through \cite{Klinkhamer:2023avf, Fuller:1962zza}. In 1988, Morris, Thorne, and Yurtsever explicitly showed a way to transform a wormhole from a purely spatial passage into a conduit for time travel \cite{morris1988, Morris:1988tu}. Following this, other forms of traversable wormholes were identified as valid solutions to Einstein’s equations. Notably, Visser's 1989 study introduced thin-shell wormholes, where a traversing path could avoid regions of exotic matter altogether \cite{visser1996}. Despite this, the presence of exotic matter remains a significant challenge for constructing wormholes. More recently, wormholes have also been proposed as a key concept in understanding quantum entanglement \cite{Marolf:2013dba}.
\vspace{0.10cm}

\setlength{\parindent}{0pt}

In this work, we consider the stationary, axially symmetric rotating Teo wormhole \cite{Teo:1998dp}, the first rotating wormhole solution and a generalization of the Morris-Thorne model \cite{morris1988, Morris:1988tu}. Although it provides a fascinating theoretical framework, it violates the null energy condition \cite{Tsukamoto:2014swa}, and detecting such wormholes remains a major challenge. Light deflection by wormholes was first noted for the Ellis wormhole by Chetouani and Clément \cite{Chetouani:1984qdm}, and since then, both weak and strong lensing effects have been widely studied. Tsukamoto examined deflection limits in Ellis spacetime \cite{Tsukamoto:2016zdu, Tsukamoto:2012zz, Tsukamoto:2012xs}, while Nakajima and Asada explored gravitational lensing \cite{Nakajima:2012pu}, and Bhattachary and Potapov applied various analytical methods to compute the deflection angle \cite{Bhattacharya:2010zzb}. Further studies have addressed microlensing and retro-lensing \cite{Abe:2010ap, Tsukamoto:2017edq}, as well as strong deflection in Janis–Newman–Winnicour and Ellis wormholes \cite{Tsukamoto:2016qro, Dey:2008kn}. Other works have investigated lensing in brane-world and scalar-tensor wormholes \cite{Nandi:2006ds, Shaikh:2017zfl}, wave effects in lensing \cite{Yoo:2013cia}, the formation of primordial wormholes in the early universe \cite{Nojiri:1999vv, Nojiri:1999pc}, and frame-dragging and light deflection in rotating optical wormhole spacetimes \cite{Errehymy:2025psi}.
\vspace{0.10cm}

\setlength{\parindent}{0pt}

While most previous studies focused on non-rotating wormholes, rotating systems are more relevant from an astrophysical standpoint. Motivated by this, we shall explore slowly rotating traversable wormholes embedded in realistic galactic dark matter halos. In the early 1970s, astronomers realized that the visible matter in certain galaxies was insufficient to generate the gravitational pull needed to explain the high rotation speeds observed in their outer regions \cite{Freeman:1970mx, Whitehurst:1972, Rogstad:1972, Roberts:1973}. This deficit in visible matter led to the hypothesis of dark matter---an unseen, non-luminous component---which has since been confirmed in many galaxies \cite{Rubin:1978kmz, deBlok:2002a, Walter:2008, Lelli:2016nwa}. At present, it is generally recognized that a significant fraction of the mass in galaxies---and possibly the Universe as a whole---exists in the form of dark matter. While alternative explanations, such as modifications to Newtonian dynamics \cite{Milgrom:1983pn, Begeman:1991, Sanders:2007, Swaters:2010} or Newtonian gravity \cite{Brownstein:2005zz, Cardone:2010, Lin:2013}, can partially address the observed discrepancies, the dark matter framework remains the most successful in explaining key phenomena, including galaxy formation and the cosmic microwave background radiation.
\vspace{0.10cm}

\setlength{\parindent}{0pt}

Mapping the distribution of dark matter in galaxies is essential, as it governs both galactic dynamics and cosmic evolution. Several dark matter profiles have been proposed: the NFW model by Navarro et al. \cite{Navarro:1995iw, Navarro:1996gj} from N-body simulations in cold dark matter cosmology, the Burkert profile \cite{burkert1995} for dwarf spiral galaxies, and the pseudo-isothermal profile by Jimenez et al. \cite{jimenez2003} fitting a wide range of galaxy rotation curves. Brownstein \cite{Brownstein:2009} further introduced a core-modified profile with constant central density that reproduces rotation curves of both high- and low-surface brightness galaxies. These models generally involve two parameters, the characteristic density $\rho_s$ and scale length $R_s$, while other profiles, such as the Einasto \cite{einasto1965}  or generalized profiles \cite{Zhao:1995cp, An:2012pv} , include additional parameters to capture more complex halo structures. 
\vspace{0.10cm}

\setlength{\parindent}{0pt}

Here we are going to carry out a comprehensive study of slowly rotating traversable wormholes embedded within realistic galactic dark matter halos, concentrating on three widely used density profiles: Navarro-Frenk-White (NFW \cite{Navarro:1995iw, Navarro:1996gj}), Bose-Einstein Condensate (Thomas-Fermi, TF)~\cite{Boehmer:2007um}, and pseudo-isothermal (PI) ~\cite{Begeman:1991iy} models. By utilizing the Teo-type rotating wormhole metric, we shall construct the corresponding shape functions directly from these halo density distributions, which enables a detailed analysis of the geometrical features of the wormholes. In particular, we shall examine the structure of the wormhole throat, the satisfaction of the flaring-out condition necessary for traversability, and the associated violations of the null energy condition, which are essential for understanding the physical feasibility of these configurations. Beyond the purely geometrical aspects, we shall investigate the motion of photons in these spacetimes by analyzing null geodesics, effective potentials, photon spheres, and LT precession caused by the rotation of the wormholes. This analysis allows us to explore how different halo structures---cuspy profiles such as NFW versus cored profiles like TF and PI---affect photon trajectories and the dynamics of test particles around the wormholes. The influence of the halo type on relativistic effects, such as frame dragging and orbit stability, is expected to produce distinct observational signatures, which can help differentiate between different dark matter environments. Finally, we shall compute the shadows cast by these wormholes, establishing a direct link between their theoretical geometry and potential observational effects. We shall demonstrate that cuspy NFW halos tend to generate smaller, asymmetric shadows, whereas cored TF and PI halos lead to smoother, nearly circular silhouettes. By combining the study of wormhole geometry, energy condition violations, photon dynamics, and shadow morphology, our work provides a thorough theoretical framework for characterizing slowly rotating traversable wormholes in diverse galactic dark matter halos. This analysis not only enhances our understanding of the interplay between dark matter distributions and wormhole properties but also offers a foundation for identifying potential observational signatures of these exotic objects in realistic astrophysical environments.

\section{Rotating Wormhole Metric Tensor}\label{sec:2}

In this section, we examine the spacetime geometry associated with a rotating wormhole configuration. As discussed in~\cite{Teo:1998dp}, such a metric is both stationary and axisymmetric, admitting a timelike Killing vector field 
\(\zeta^{a} \equiv (\partial / \partial t)^{a}\), corresponding to invariance under time translations, and a spacelike Killing vector field 
\(\psi^{a} \equiv (\partial / \partial \varphi)^{a}\), corresponding to invariance under rotations about the azimuthal axis. 
According to~\cite{Papapetrou, Carter1, Carter1a, Carter2}, the most general stationary and axisymmetric line element may be written as
\begin{align}
ds^{2} = g_{tt} \, dt^{2} + 2 g_{t\varphi} \, dt \, d\varphi 
+ g_{\varphi\varphi} \, d\varphi^{2} + g_{ij} \, dx^{i} dx^{j},
\label{metric_general1}
\end{align}
where $i,j$ denote the remaining spatial coordinates. 

It has been suggested that introducing a time-dependent conformal factor into the Morris-Thorne wormhole metric can partially alleviate violations of the null energy condition (NEC). However, such a modification leads to an isotropically expanding throat, thereby destroying the stationary character of the wormhole and rendering it unsuitable for practical traversability~\cite{Teo:1998dp, Roman:1993, Kar:1994, Kar:1996}.

\vspace{0.1cm}
\setlength{\parindent}{0pt}

\begin{figure}[!htbp]
\centering 
\includegraphics[scale=0.99]{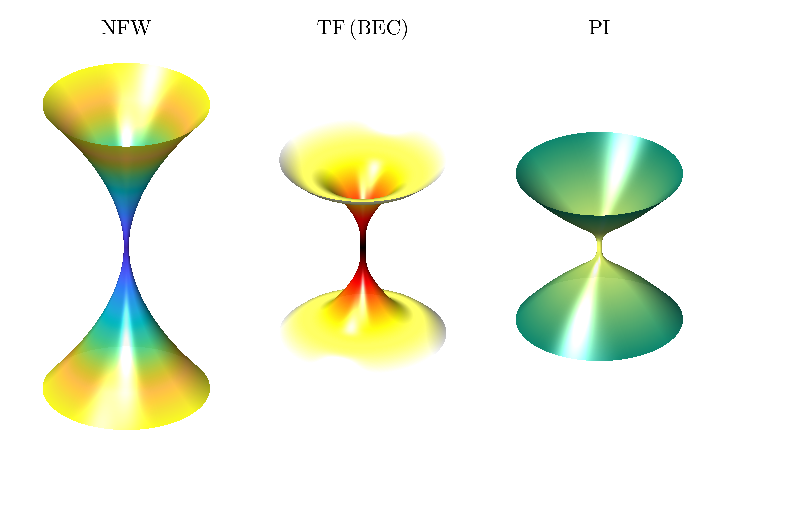} 
\caption{\fontsize{8}{9}\selectfont Three-dimensional embedding diagrams of static, spherically symmetric wormholes sustained by different dark matter halo profiles. From left to right, the panels correspond to the NFW, TF (BEC), and PI halos. Each panel depicts the upper ($z>0$) and lower ($z<0$) universes, smoothly joined at the throat radius $r_0 = 0.2$. The rotationally symmetric surfaces are generated using $r \in [r_0, r_{\rm max}]$ and $\phi \in [0, 2\pi]$. Characteristic parameters are: $\rho_s^{\rm NFW} = 0.02$, $R_s^{\rm NFW} = 1.0$; $\rho_s^{\rm TF} = 0.02$, $R_s^{\rm TF} = 1.0$; $\rho_s^{\rm PI} = 0.02$, $R_s^{\rm PI} = 1.0$. The embedding functions are computed numerically to illustrate geometrical differences induced by each halo profile.} 
\label{fig:3D} 
\end{figure}

In the present work, we focus on a rotating wormhole spacetime expressed in spherical polar coordinates, as introduced in~\cite{Teo:1998dp}:
\begin{align}
ds^{2} = - e^{2\Phi(r)} dt^{2} 
+ \frac{dr^{2}}{1 - \dfrac{\epsilon(r)}{r}} 
+ r^{2} K(r)^{2} \left[ d\theta^{2} 
+ \sin^{2}\theta \, ( d\varphi - \omega(r) dt )^{2} \right],
\label{metric_general2}
\end{align}
where $\Phi(r)$ is the redshift function, finite everywhere to prevent event horizon formation and ensure traversability~\cite{Butcher:2015sea}, and $\epsilon(r)$ is the shape function specifying the spatial profile of the wormhole. For different shape functions, the static wormhole surfaces are shown in Figure~\ref{fig:3D}.

The function $\omega(r)$ denotes the angular velocity of the rotating configuration. The radial coordinate $r$ achieves its minimum value at the throat $r_{0}$, where 
\(\epsilon(r_{0}) = r_{0}\). Near the throat, the geometry satisfies the flaring-out condition~\cite{Morris:1988cz}:
\begin{align}
\epsilon(r) - r \, \epsilon'(r) \geq 0, 
\end{align}
where the prime denotes differentiation with respect to $r$. Additionally, one requires $\epsilon(r)/r \to 0$ as $r \to \infty$, with $\epsilon(r)/r < 1$ everywhere to avoid horizons. 

The function $K(r)$ is positive and non-decreasing, describing the proper radial distance as a function of $r$. For slowly rotating wormholes, one often assumes $K(r) \approx 1$. This metric form was previously applied by Hartle~\cite{Hartle:1967ha, Hartle:1967he} in studies of relativistic rotating stars. To guarantee asymptotic flatness, the metric functions must satisfy
\begin{align}
\Phi(r) \to 0, \qquad K(r) \to 1, \qquad \omega(r) \to 0 \quad (r \to \infty).
\label{asymptotic_condition}
\end{align}
The angular velocity function is typically assumed to decay as
\begin{align}
\omega(r) = \frac{2 J}{r^{3}} + \mathcal{O}(r^{-4}),
\end{align}
where $J$ denotes the total angular momentum of the configuration~\cite{Teo:1998dp}. In the linear (slow rotation) approximation, the contribution of $\omega(r)$ to the NEC and the shape function is negligible.

\section{Commonly Considered Models for Galactic Dark Matter Halos} \label{sec:3}

Several studies have proposed that certain galaxies may host traversable wormholes at their centers~\cite{Kuhfittig:2013hva, Rahaman:2013xoa, Rahaman:2014pba}, with much of the supporting evidence inferred from galactic rotation curves. 
The concept of dark matter originated as an explanation for the nearly flat rotation curves of spiral galaxies, dating back to the measurement of the Oort constants~\cite{Oort:1927a} and Zwicky’s pioneering mass estimates of galaxy clusters~\cite{Zwicky:1979a, Zwicky:1979b}. 

To reproduce these curves, the Navarro-Frenk-White (NFW) profile was introduced~\cite{Navarro:1995iw, Navarro:1996gj}, providing excellent agreement with large-scale cosmological $N$-body simulations~\cite{Graham:2005xx}. 
In a galactic wormhole scenario, the wormhole throat is typically embedded in a dark matter halo, with the halo’s density profile determining the overall geometry. 
Violations of classical energy conditions near the throat may be interpreted as indirect evidence of the halo’s gravitational influence. 
In this section, we review three commonly considered halo density profiles and their implications for wormhole geometries.

\subsection{The Navarro-Frenk-White (NFW) Profile}\label{sec:3:1}

Within the Cold Dark Matter (CDM) paradigm, the NFW profile remains one of the most widely used descriptions of halo structure, derived from high-resolution $N$-body simulations~\cite{Jusufi:2019knb, Xu:2020wfm, Dubinski:1991bm, Navarro:1995iw, Navarro:1996gj}. 
Its density distribution is
\begin{equation}
\label{NFW}
\rho_{\mathrm{NFW}}(r) = 
\frac{\rho_{s}}{\frac{r}{R_{s}} \left( 1 + \frac{r}{R_{s}} \right)^{2}} 
= \frac{\rho_s R_s^3}{r (r + R_s)^2},
\end{equation}
where $R_s$ is the characteristic scale radius and $\rho_s$ is the density normalization. 
For the M87 galaxy, $\rho_s = 0.008 \times 10^{7.5} \, M_\odot \,\mathrm{kpc}^{-3}$~\cite{Oldham:2016a} and $R_s = 130~\mathrm{kpc}$~\cite{Jusufi:2019knb}.  

\subsection{The Thomas-Fermi (TF) Profile}\label{sec:3:2}

In the Bose-Einstein Condensate (BEC) dark matter model, the TF approximation yields a cored density profile~\cite{Boehmer:2007um}:
\begin{equation}
\label{TF}
\rho_{\mathrm{TF}}(r) = \rho_s \, \frac{\sin(\pi r / R_s)}{\pi r / R_s},
\end{equation}
where $\rho_s$ is the central density and $R_s$ is the radius at which both density and pressure vanish. 
Unlike NFW, the TF profile has a finite central density: \(\rho_{\rm TF}(0) = \rho_s\). 
For the Milky Way, typical values are~\cite{Hou:2018bar, Xu:2018wow}:
\begin{equation*}
\rho_s = 3.43 \times 10^7 \, M_\odot \,\mathrm{kpc}^{-3}, 
\qquad R_s = 15.7 \,\mathrm{kpc}.
\end{equation*}

\subsection{The Pseudo-Isothermal (PI) Profile}\label{sec:3:3}

The PI profile is particularly useful in Modified Newtonian Dynamics-inspired phenomenology~\cite{Begeman:1991iy}:
\begin{equation}
\label{PI}
\rho_{\mathrm{PI}}(r) = \frac{\rho_s}{1 + (r/R_s)^2},
\end{equation}
where $\rho_s$ is the central density and $R_s$ is the core radius. 
For ESO1200211, $\rho_s = 0.0464 \, M_\odot \,\mathrm{pc}^{-3}$ and $R_s = 0.57 \,\mathrm{kpc}$ \cite{Robles:2012uy}.

\section{Shape Functions for Dark Matter Halo Profiles and Slowly Rotating Wormholes}\label{sec:4}

The shape function $\epsilon(r)$ determines the geometry of static or slowly rotating wormholes. For a spherically symmetric spacetime sourced by density profile $\rho(r)$, the Einstein field equations give
\begin{equation}
\epsilon'(r) = 8 \pi r^2 \rho(r),
\end{equation}
where \(\epsilon'(r) \equiv d\epsilon / dr\). The integration constant \(\epsilon_0\) is fixed by the throat condition \(\epsilon(r_0) = r_0\).

The radial null energy condition (NEC) reads
\begin{equation}
\rho + p_r = \frac{1}{8 \pi r^2} \left(\epsilon' - \frac{\epsilon}{r}\right) + \frac{1 - \epsilon/r}{4 \pi r} \Phi'(r),
\end{equation}
where $\Phi(r)$ is the redshift function. Substituting the shape functions and their derivatives enables direct evaluation of the NEC, identifying regions where it is satisfied or violated. These results are presented in Table~\ref{tab:shape_nec}.

\section{Wormhole Scenarios}\label{sec:5}

In this section, we present explicit wormhole metrics obtained by substituting the shape functions $\epsilon(r)$ corresponding to the three representative dark matter halo profiles. 
This procedure illustrates how the choice of halo model influences the throat geometry, the flaring-out condition, and the amount of exotic matter required to sustain the wormhole. 
We also discuss the extension to slowly rotating configurations.

\begin{table*}[!htbp]
\centering
\scriptsize
\setlength{\tabcolsep}{5pt} 
\renewcommand{\arraystretch}{1.5} 

\rowcolors{2}{gray!15}{white}
\begin{tabular}{c p{5.cm} p{3.9cm} p{3.9cm}}
\hline\hline
\rowcolor{blue!20}
\textbf{Profile} & \textbf{$\epsilon(r)$} & \textbf{$\epsilon'(r)$} & \textbf{$\rho + p_r$} \\
\hline\hline
\textbf{NFW} & 
$\displaystyle \epsilon_{\text{NFW}}(r) = 8 \pi \rho_s R_s^3 \Big[\ln(r + R_s) + \frac{R_s}{r + R_s}\Big] + \epsilon_0$ &
$\displaystyle \epsilon'_{\text{NFW}}(r) =  8 \pi \rho_s R_s^3 \frac{r}{(r + R_s)^2}$ &
$\displaystyle \rho + p_r = \frac{1}{8 \pi r^2}\Big(\epsilon' - \frac{\epsilon}{r}\Big) + \frac{1 - \epsilon/r}{4 \pi r}\,\Phi'(r)$
\\[1ex]
\textbf{TF (BEC)} & 
$\displaystyle \epsilon_{\text{TF}}(r) = 8 \pi \rho_s R_s^3 \Big[-\frac{r}{\pi R_s}\cos\Big(\frac{\pi r}{R_s}\Big) + \frac{1}{\pi^2}\sin\Big(\frac{\pi r}{R_s}\Big)\Big] + \epsilon_0$ &
$\displaystyle \epsilon'_{\text{TF}}(r) = 8 \pi \rho_s R_s r \sin\Big(\frac{\pi r}{R_s}\Big)$ &
$\displaystyle \rho + p_r = \frac{1}{8 \pi r^2}\Big(\epsilon' - \frac{\epsilon}{r}\Big) + \frac{1 - \epsilon/r}{4 \pi r}\,\Phi'(r)$
\\[1ex]
\textbf{PI} & 
$\displaystyle \epsilon_{\text{PI}}(r) = 8 \pi \rho_s R_s^3 \Big[\frac{r}{R_s} - \arctan\!\Big(\frac{r}{R_s}\Big)\Big] + \epsilon_0$ &
$\displaystyle \epsilon'_{\text{PI}}(r) = 8 \pi \rho_s \frac{r^2}{1 + (r / R_s)^2}$ &
$\displaystyle \rho + p_r = \frac{1}{8 \pi r^2}\Big(\epsilon' - \frac{\epsilon}{r}\Big) + \frac{1 - \epsilon/r}{4 \pi r}\,\Phi'(r)$
\\
\hline\hline
\end{tabular}
\caption{\fontsize{8}{9}\selectfont Shape functions, derivatives, and radial NEC for representative dark matter halo profiles.}
\label{tab:shape_nec}
\end{table*}

\subsection{(i) NFW Halo Supported Wormhole}\label{sec:5:1}

For a wormhole embedded in a Navarro-Frenk-White (NFW) dark matter halo, the static, spherically symmetric spacetime metric reads
\begin{equation}
ds^2 = - e^{2\Phi(r)} dt^2 
+ \frac{dr^2}{1 - \dfrac{\epsilon_{\mathrm{NFW}}(r)}{r}}
+ r^2 \left(d\theta^2 + \sin^2\theta \, d\varphi^2\right),
\end{equation}
with the NFW shape function
\begin{equation}
\epsilon_{\mathrm{NFW}}(r) = 8 \pi \, \rho_s R_s^3 \left[ \ln(r+R_s) + \frac{R_s}{r+R_s} \right] + \epsilon_0,
\end{equation}
where $\rho_s$ and $R_s$ are the characteristic density and scale radius of the halo, and $\epsilon_0$ is fixed by the throat condition $\epsilon(r_0)=r_0$. The NFW profile is cuspy at the center, producing a sharper throat geometry and typically stronger violations of the null energy condition (NEC) near the throat. The metric remains asymptotically flat as $r \to \infty$, satisfying $\epsilon(r)/r \to 0$ and $\Phi(r) \to 0$. 

For a slowly rotating wormhole, the metric generalizes to include frame-dragging:
\begin{equation}
ds^2 = - e^{2\Phi(r)} dt^2 
+ \frac{dr^2}{1 - \epsilon_{\mathrm{NFW}}(r)/r} 
+ r^2 \left[d\theta^2 + \sin^2\theta \left(d\varphi - \omega(r) dt\right)^2\right],
\end{equation}
where $\omega(r) \simeq 2 J / r^3$ is the angular velocity, $J$ being the total angular momentum. Slow rotation mainly affects frame-dragging and LT precession near the throat, leaving the radial structure governed by the shape function.

\subsection{(ii) Thomas-Fermi (BEC) Halo Supported Wormhole} \label{sec:5:2}

For a wormhole sourced by a Bose-Einstein Condensate (BEC) or Thomas-Fermi (TF) dark matter halo:
\begin{equation}
ds^2 = - e^{2\Phi(r)} dt^2 
+ \frac{dr^2}{1 - \dfrac{\epsilon_{\mathrm{TF}}(r)}{r}}
+ r^2 \left(d\theta^2 + \sin^2\theta \, d\varphi^2\right),
\end{equation}
with the shape function
\begin{equation}
\epsilon_{\mathrm{TF}}(r) = 8 \pi \, \rho_s R_s^3 \left[
- \frac{r}{\pi R_s} \cos\left(\frac{\pi r}{R_s}\right)
+ \frac{1}{\pi^2} \sin\left(\frac{\pi r}{R_s}\right)
\right] + \epsilon_0.
\end{equation}
The TF halo is cored, producing a smoother throat and requiring less exotic matter. Slow rotation modifies the metric as
\begin{equation}
ds^2 = - e^{2\Phi(r)} dt^2 
+ \frac{dr^2}{1 - \epsilon_{\mathrm{TF}}(r)/r} 
+ r^2 \left[d\theta^2 + \sin^2\theta \left(d\varphi - \omega(r) dt\right)^2\right].
\end{equation}

\subsection{(iii) Pseudo-Isothermal (PI) Halo Supported Wormhole}\label{sec:5:3}

For a pseudo-isothermal (PI) halo:
\begin{equation}
ds^2 = - e^{2\Phi(r)} dt^2 
+ \frac{dr^2}{1 - \dfrac{\epsilon_{\mathrm{PI}}(r)}{r}}
+ r^2 \left(d\theta^2 + \sin^2\theta \, d\varphi^2\right),
\end{equation}
with shape function
\begin{equation}
\epsilon_{\mathrm{PI}}(r) = 8 \pi \, \rho_s R_s^3 \left[ \frac{r}{R_s} - \arctan\left(\frac{r}{R_s}\right) \right] + \epsilon_0.
\end{equation}
Cored profiles like PI and TF favor smoother throats and smaller NEC violations. The slowly rotating generalization is
\begin{equation}
ds^2 = - e^{2\Phi(r)} dt^2 
+ \frac{dr^2}{1 - \epsilon_{\mathrm{PI}}(r)/r} 
+ r^2 \left[d\theta^2 + \sin^2\theta \left(d\varphi - \omega(r) dt\right)^2\right].
\end{equation}

\section{Null Geodesics and Photon Dynamics in Slowly Rotating Dark Matter Halo Wormholes}\label{sec:6}

We now analyze null geodesics in slowly rotating wormholes supported by NFW, TF, and PI halos. Slow rotation introduces LT precession through $g_{t\varphi} = - r^2 \omega(r)$, modifying photon trajectories.

\subsection{Lense-Thirring Precession Near the Throat}\label{sec:6:1}

The LT precession frequency in the equatorial plane ($\theta = \pi/2$), to first order in $\omega(r)$, is
\begin{equation}
\Omega_\mathrm{LT}(r) = -\frac{1}{2} \frac{d\omega(r)}{dr}.
\end{equation}
For $\omega(r) \simeq 2 J / r^3$, we obtain
\begin{equation}
\Omega_\mathrm{LT}(r) = \frac{3 J}{r^4}.
\end{equation}

Representative values near typical throat radii:
\begin{itemize}
    \item \textbf{NFW cusp:} $r_0 = 0.2$, $\Omega_\mathrm{LT}^\mathrm{NFW} \approx 93.8$,
    \item \textbf{TF core:} $r_0 = 0.5$, $\Omega_\mathrm{LT}^\mathrm{TF} \approx 2.4$,
    \item \textbf{PI core:} $r_0 = 0.5$, $\Omega_\mathrm{LT}^\mathrm{PI} \approx 2.4$.
\end{itemize}

\subsection{Stationary, Axisymmetric Metric and Null Geodesics}\label{sec:6:2}

Using the Teo-type metric:
\begin{equation}
ds^{2} = - e^{2\Phi(r)} dt^{2} + \frac{dr^{2}}{1 - \epsilon(r)/r} + r^{2} \Big[ d\theta^{2} + \sin^{2}\theta \left( d\varphi - \omega(r) dt \right)^{2} \Big],
\end{equation}
and restricting to $\theta = \pi/2$:
\begin{equation}
ds^{2} = - e^{2\Phi(r)} dt^{2} + \frac{dr^{2}}{1 - \epsilon(r)/r} + r^{2} (d\varphi - \omega(r) dt)^2.
\end{equation}

The null condition $ds^2 = 0$ leads to conserved quantities:
\begin{align}
E &= e^{2\Phi(r)} \dot{t} + r^2 \omega(r) \dot{\varphi}, \\
L &= r^2 (\dot{\varphi} - \omega(r) \dot{t}),
\end{align}
with
\begin{align}
\dot{t} &= \frac{E - \omega(r) L}{e^{2\Phi(r)}}, \\
\dot{\varphi} &= \frac{L}{r^2} + \omega(r) \frac{E}{e^{2\Phi(r)}}.
\end{align}

The radial equation becomes
\begin{equation}
\dot{r}^2 = \left(1 - \frac{\epsilon(r)}{r}\right) \left[ \frac{(E - \omega(r) L)^2}{e^{2\Phi(r)}} - \frac{L^2}{r^2} \right].
\end{equation}

Defining the effective potential
\begin{equation}
V_\mathrm{eff}(r) = - \left(1 - \frac{\epsilon(r)}{r}\right) \left[ \frac{(E - \omega(r) L)^2}{e^{2\Phi(r)}} - \frac{L^2}{r^2} \right],
\end{equation}
radial motion satisfies $\dot{r}^2 + V_\mathrm{eff}(r) = 0$. Local maxima of $V_\mathrm{eff}(r)$ define photon spheres.

The angular trajectory is
\begin{equation}
\frac{d\varphi}{dr} = \frac{\frac{L}{r^2} + \omega(r) \frac{E}{e^{2\Phi(r)}}}{\sqrt{\left(1 - \frac{\epsilon(r)}{r}\right) \left[ \frac{(E - \omega(r) L)^2}{e^{2\Phi(r)}} - \frac{L^2}{r^2} \right]}}.
\end{equation}

Introducing $b = L/E$ and linearizing in $\omega(r)$:
\begin{equation}
\varphi(r) = \int \frac{\frac{b}{r^2} + \frac{\omega(r)}{e^{2\Phi(r)}}}{\sqrt{\left(1 - \frac{\epsilon(r)}{r}\right) \left( e^{-2\Phi(r)} - \frac{b^2}{r^2} \right)}} \, dr + \mathcal{O}(\omega^2),
\end{equation}
so the total deflection angle is
\begin{equation}
\hat{\alpha} = 2 \int_{r_\mathrm{min}}^\infty \frac{d\varphi}{dr} dr - \pi.
\end{equation}

Co-rotating photons ($b>0$) experience slightly reduced deflection, while counter-rotating photons ($b<0$) experience enhanced deflection.

\subsection{Photon Spheres and Rotational Splitting}\label{sec:6:3}

Circular null orbits satisfy $\dot{r} = 0$ and $\ddot{r} = 0$, leading to
\begin{equation}
\frac{(E - \omega(r) L)^2}{e^{2\Phi(r)}} - \frac{L^2}{r^2} = 0, \qquad
\frac{d}{dr} \left[ \frac{(E - \omega(r) L)^2}{e^{2\Phi(r)}} - \frac{L^2}{r^2} \right] = 0.
\end{equation}

Introducing the impact parameter $b = L/E$ and expanding to first order in $\omega(r)$:
\begin{equation}
\frac{1 - 2 \omega(r) b}{e^{2\Phi(r)}} \approx \frac{b^2}{r^2}, \qquad \Rightarrow \quad b_\pm(r_\mathrm{ph}) \simeq \frac{r_\mathrm{ph}}{e^{\Phi(r_\mathrm{ph})}} \left( 1 \pm \frac{\omega(r_\mathrm{ph}) r_\mathrm{ph}}{e^{\Phi(r_\mathrm{ph})}} \right),
\end{equation}
where $+$ ($-$) corresponds to counter-rotating (co-rotating) orbits. In the static limit $\omega \to 0$, we recover
\begin{equation}
b_\mathrm{ph} = \frac{r_\mathrm{ph}}{e^{\Phi(r_\mathrm{ph})}}.
\end{equation}

The shift in the photon sphere radius due to slow rotation can be estimated by linearizing around the static radius $r_\mathrm{ph}^{(0)}$:
\begin{equation}
r_\mathrm{ph} \simeq r_\mathrm{ph}^{(0)} + \delta r_\mathrm{ph}, \qquad
\delta r_\mathrm{ph} \simeq \frac{2 \, r_\mathrm{ph}^{(0)3} \, \omega(r_\mathrm{ph}^{(0)})}{2 - r_\mathrm{ph}^{(0)} \Phi'(r_\mathrm{ph}^{(0)})}.
\end{equation}

Co-rotating photons ($\omega>0$, $b>0$) are shifted inward ($\delta r_\mathrm{ph} < 0$), while counter-rotating photons are shifted outward ($\delta r_\mathrm{ph} > 0$). For representative halo models:
\begin{align}
&\delta r_\mathrm{ph}^{\mathrm{NFW}} \simeq \frac{4 J}{2 + a/r_\mathrm{ph}^{(0)}}, \quad \Phi_\mathrm{NFW}(r) = -\frac{a}{r}, \\
&\delta r_\mathrm{ph}^{\mathrm{TF}} \simeq \frac{4 J}{2 - r_\mathrm{ph}^{(0)} \Phi_\mathrm{TF}'(r_\mathrm{ph}^{(0)})}, \quad \Phi_\mathrm{TF}(r) = \ln \left[1 + \frac{A}{1 + (r/r_0)^n} \right], \\
&\delta r_\mathrm{ph}^{\mathrm{PI}} \simeq \frac{4 J}{2 - r_\mathrm{ph}^{(0)} \Phi_\mathrm{PI}'(r_\mathrm{ph}^{(0)})}, \quad \Phi_\mathrm{PI}(r) = \ln \left[1 + \frac{A}{1 + (r/r_0)^2} \right].
\end{align}

\begin{figure*}[!ht]
\centering
\includegraphics[scale=0.5]{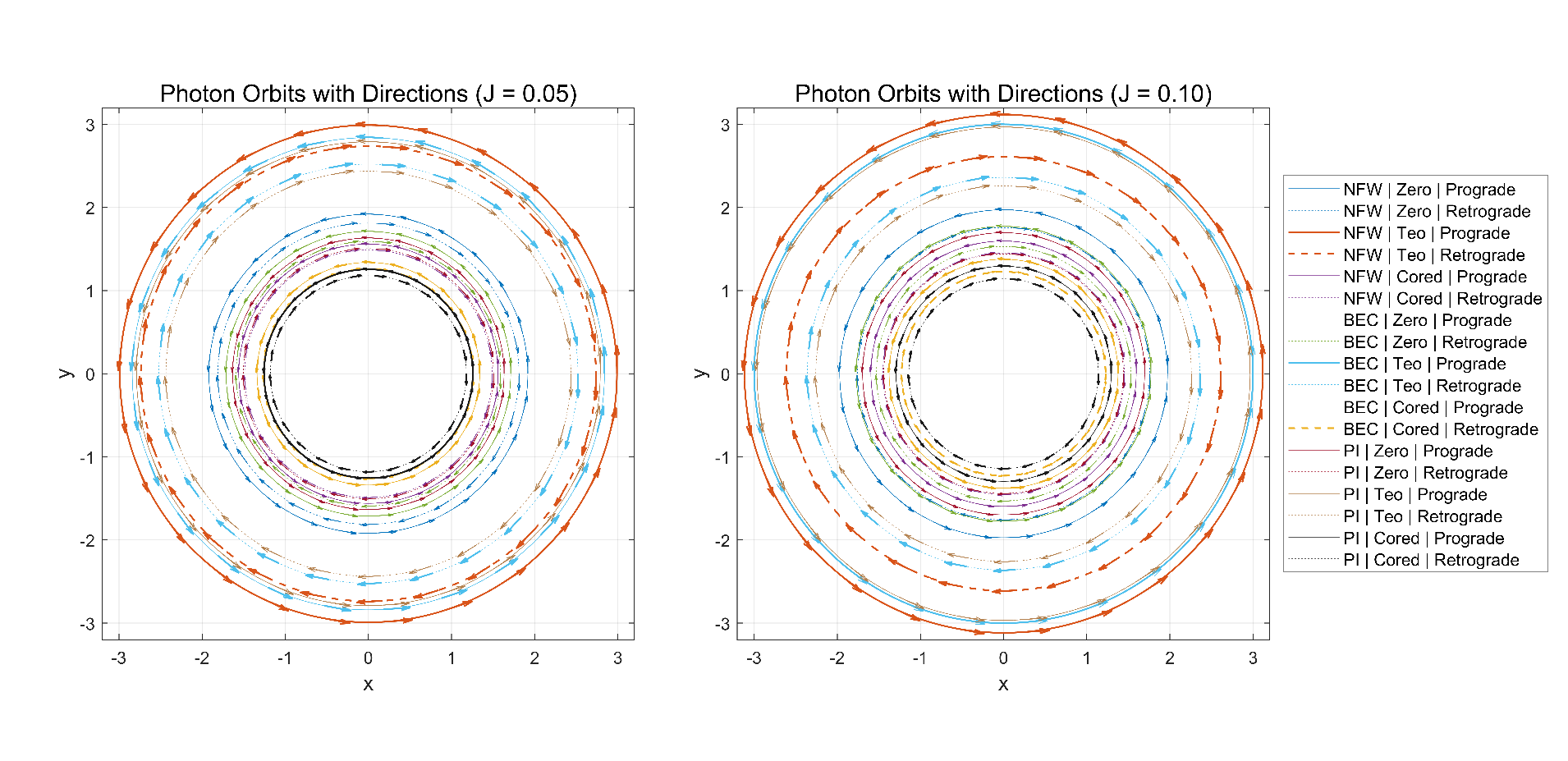}
\caption{\fontsize{8}{9}\selectfont Photon orbits of rotating wormholes embedded in three distinct dark matter halos (NFW, BEC, and PI) for rotation parameters $J = 0.05$ and $J = 0.1$. Solid lines represent prograde (co-rotating) photon orbits, dashed lines represent retrograde (counter-rotating) orbits, with arrows indicating the motion direction. Different redshift profiles, Zero ($\Phi = 0$), Teo ($\Phi(r) = -0.8/r$), and Cored ($\Phi(r) = \ln[1 + A/(1 + (r/r_0)^2)], A = 1.0$), modify the apparent photon-sphere radii. Each halo-redshift combination is color-coded, illustrating how halo type, rotation, and gravitational potential combine to shape the apparent photon-sphere.}
\label{fig:photon-orbits-direction}
\end{figure*}

The photon sphere radius and co-/counter-rotating splitting determine the wormhole shadow. Cuspy halos (NFW) produce stronger frame-dragging and sharper throat geometries, leading to asymmetric shadows. Cored halos (TF, PI) yield smoother throats with weaker frame-dragging, producing nearly circular, less distorted shadows.

Explicitly, the co-rotating ($-$) and counter-rotating ($+$) photon impact parameters satisfy
\begin{equation}
b_\pm(r_\mathrm{ph}) = \frac{r_\mathrm{ph}}{e^{\Phi(r_\mathrm{ph})}} 
\frac{1}{1 \mp \dfrac{\omega(r_\mathrm{ph}) \, r_\mathrm{ph}}{e^{\Phi(r_\mathrm{ph})}}} 
\simeq \frac{r_\mathrm{ph}}{e^{\Phi(r_\mathrm{ph})}} \left( 1 \pm \frac{\omega(r_\mathrm{ph}) \, r_\mathrm{ph}}{e^{\Phi(r_\mathrm{ph})}} \right),
\end{equation}
where the $\pm$ signs now consistently indicate counter-rotating and co-rotating trajectories, respectively. In the static limit $\omega(r_\mathrm{ph}) \to 0$, the impact parameters coincide:
\begin{equation}
b_\mathrm{ph} = \frac{r_\mathrm{ph}}{e^{\Phi(r_\mathrm{ph})}}.
\end{equation}

Hence, the dark matter halo profile affects the shadow via both the photon sphere radius $r_\mathrm{ph}$ and the frame-dragging $\omega(r_\mathrm{ph})$. Cuspy NFW halos produce stronger splitting ($b_+ - b_-$), while cored TF and PI halos reduce this asymmetry, potentially constraining the halo profile and wormhole rotation observationally.

\section{Influence of Redshift Functions on Photon Trajectories, Circular Orbits, and Wormhole Shadows}\label{sec:7}

The redshift function $\Phi(r)$ governs gravitational time dilation, the effective potential for null geodesics, and consequently photon trajectories, circular orbits, and the wormhole shadow as seen by distant observers. We consider three representative cases: an exponential Teo-type redshift function, a smooth cored-halo inspired profile, and the trivial zero-redshift case $\Phi(r)=0$.

Photon motion in a stationary, axisymmetric, slowly rotating wormhole is governed by
\begin{equation}
ds^2 = - e^{2\Phi(r)} dt^2 + \frac{dr^2}{1 - \epsilon(r)/r} 
+ r^2 \bigl(d\varphi - \omega(r) dt\bigr)^2,
\end{equation}
where $\epsilon(r)$ is the shape function and $\omega(r)$ is the slow-rotation frame-dragging. Restricting to the equatorial plane, the conserved energy $E$ and angular momentum $L$ satisfy
\begin{equation}
E = -p_t = e^{2\Phi}\dot{t} + r^2 \omega(r) \dot{\varphi}, 
\qquad
L = p_\varphi = r^2 \dot{\varphi} - r^2 \omega(r) \dot{t}.
\end{equation}

Solving this $2\times 2$ system and imposing the null condition $ds^2=0$ yields
\begin{equation}
\dot{r}^2 = \left(1-\frac{\epsilon(r)}{r}\right)
\left[\frac{(E - \omega(r) L)^2}{e^{2\Phi(r)}} - \frac{L^2}{r^2} \right].
\end{equation}

The orbit equation reads
\begin{equation}
\frac{d\varphi}{dr} = \frac{\displaystyle \frac{L}{r^2} + \omega(r)\frac{E - \omega(r) L}{e^{2\Phi(r)}}}
{\displaystyle \sqrt{\left(1-\frac{\epsilon(r)}{r}\right) 
\left[\frac{(E - \omega(r) L)^2}{e^{2\Phi(r)}} - \frac{L^2}{r^2} \right]}}.
\end{equation}

\begin{figure*}[ht]
\centering
\includegraphics[scale=0.5]{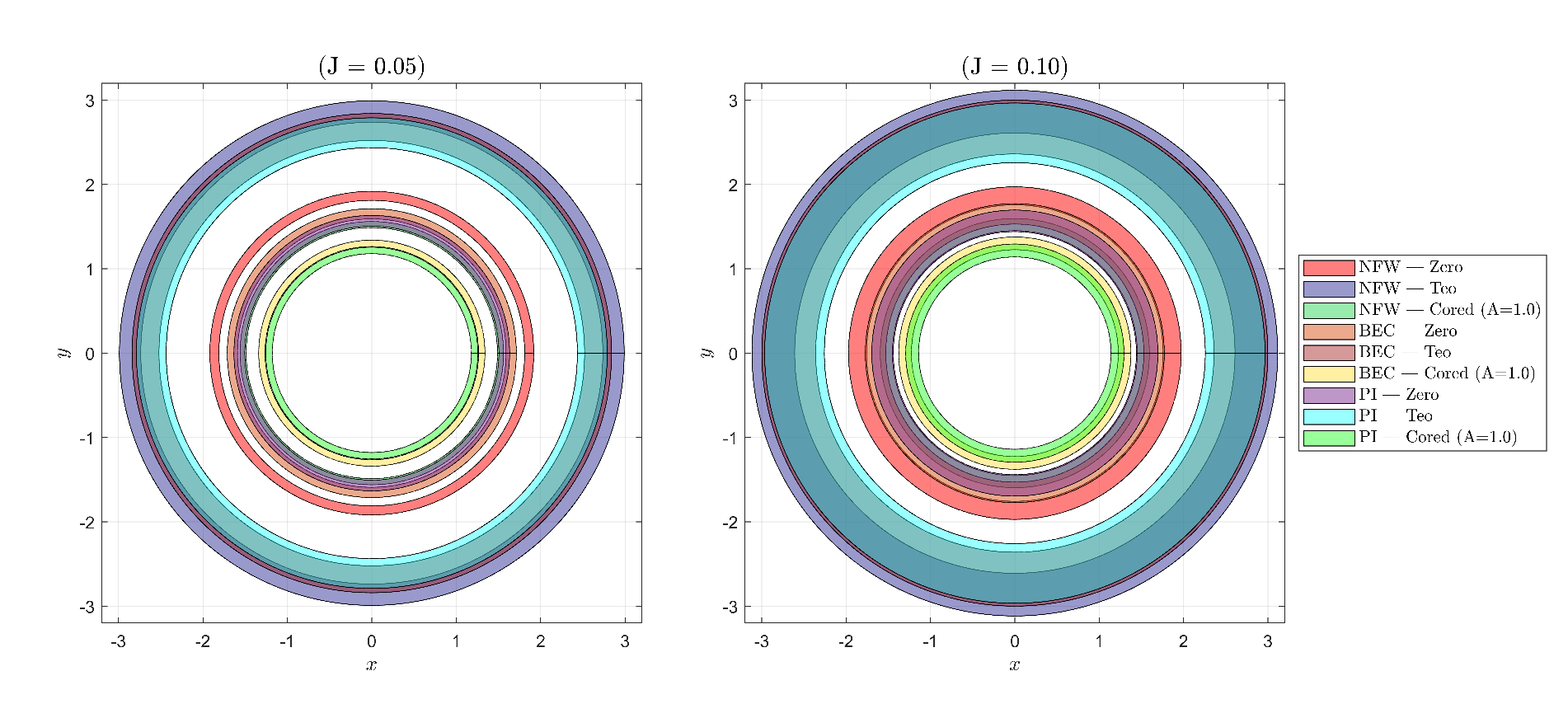}
\caption{\fontsize{8}{9}\selectfont Photon-sphere profiles of rotating wormholes embedded in three distinct dark matter halos for rotation parameters $J = 0.05$ and $J = 0.1$. The halos are NFW ($\rho_s = 0.2, r_s = 1.0$), BEC ($\rho_c = 0.1, r_c = 1.5$), and PI ($\rho_0 = 0.08, r_c = 2.0$), including the corresponding photon-sphere shifts. Three redshift profiles are shown: Zero ($\Phi = 0$), Teo ($\Phi(r) = -0.8/r$), and Cored ($\Phi(r) = \log[1 + A/(1 + (r/r_0)^2)]$, $A = 1$). Each halo-redshift combination is represented with a distinct color, illustrating how rotation and halo properties affect the photon-sphere structure and induce asymmetries in co- and counter-rotating trajectories.}
\label{fig:photon-spheres}
\end{figure*}

Under $|\omega r| \ll 1$, terms $\mathcal{O}(\omega^2)$ can be neglected. Defining $b=L/E$:
\begin{equation}
\varphi(r) = \int \frac{\displaystyle \frac{b}{r^2} + \frac{\omega(r)}{e^{2\Phi(r)}}}{\displaystyle \sqrt{\left(1 - \frac{\epsilon(r)}{r}\right) \left[ \frac{(1 - \omega(r)b)^2}{e^{2\Phi(r)}} - \frac{b^2}{r^2} \right] }} \, dr,
\end{equation}
with the positive branch of $\dot{r}$ corresponding to photons moving outward from closest approach. In the static limit $\omega(r)\to 0$, this reduces to
\begin{equation}
\varphi(r) = \int \frac{b/r^2}{\sqrt{\left(1 - \frac{\epsilon(r)}{r}\right) \left(e^{-2\Phi(r)} - b^2/r^2\right)}} \, dr,
\end{equation}
showing that even nonrotating wormholes produce deflection sensitive to $\Phi(r)$. Traversability requires $1-\epsilon(r)/r>0$, and real photon trajectories require $b < r e^{-\Phi(r)}$.

\begin{figure*}[ht]
\centering
\includegraphics[scale=0.57]{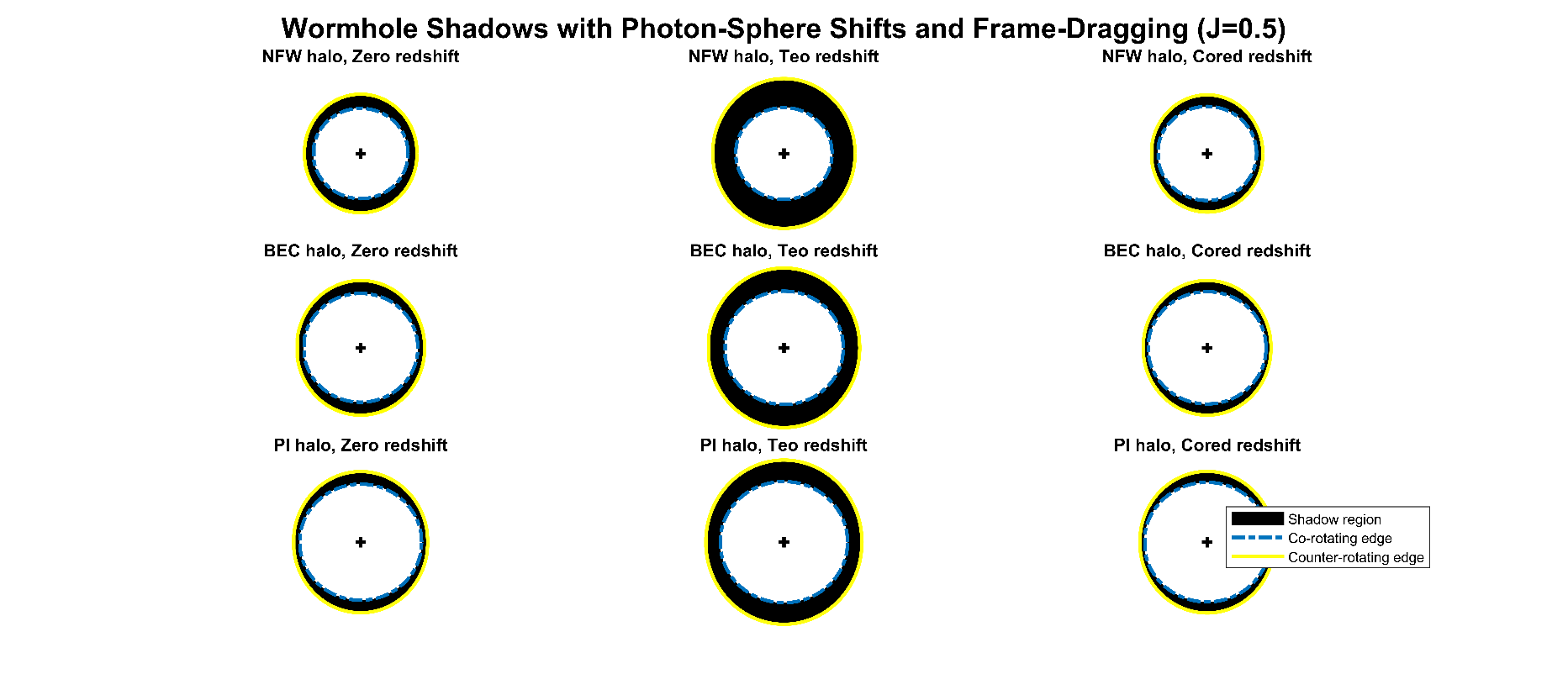}
\caption{\fontsize{8}{9}\selectfont Wormhole shadow profiles for three different dark matter halo models: NFW, BEC, and PI, under three distinct redshift profiles: Zero ($\Phi = 0$), Teo-type ($\Phi(r) = -0.8/r$), and Cored ($\Phi(r) = \log[1 + 1/(1 + (r/1.0)^2)]$), with a slow rotation parameter $J = 0.5$. The static photon-sphere radii are $r_\mathrm{ph}^\mathrm{NFW} = 3.0$, $r_\mathrm{ph}^\mathrm{BEC} = 3.5$, and $r_\mathrm{ph}^\mathrm{PI} = 3.7$, and the effective shape function approximations are $\epsilon_\mathrm{eff}^\mathrm{NFW} = 0.5$, $\epsilon_\mathrm{eff}^\mathrm{BEC} = 0.7$, $\epsilon_\mathrm{eff}^\mathrm{PI} = 0.8$. Each subplot shows the photon-sphere shift $\delta r$ computed from $\delta r = -r_0^3 \Phi'(r_0)/[2(r_0 - \epsilon)]$, producing realistic co-rotating ($b_-$, blue-dashed inner boundary) and counter-rotating ($b_+$, yellow-solid outer boundary) impact parameters. The shaded black annular region represents the wormhole shadow, with slight ellipticity to reflect frame-dragging asymmetry. The wormhole center is indicated by a black cross. Subplots are arranged in a 3$\times$3 grid: rows correspond to halos (NFW, BEC, PI), and columns correspond to redshift profiles (Zero, Teo, Cored).}
\label{fig:shadow}
\end{figure*}

\begin{table*}[!htbp]
\centering
\footnotesize
\renewcommand{\arraystretch}{1.5}
\setlength{\tabcolsep}{4pt}

\begin{tabular}{>{\raggedright}p{1.2cm} >{\raggedright}p{3cm} >{\centering}p{1.2cm} >{\centering}p{2.5cm} >{\raggedright}p{2.7cm} >{\raggedright\arraybackslash}p{3.7cm}}\hline\hline
\rowcolor{blue!10}
\textbf{Halo Profile} & \textbf{Redshift Function $\Phi(r)$} & \textbf{$\omega(r)$} & \textbf{Photon Sphere $r_\mathrm{ph}$} & \textbf{Co-/Counter-Rotating Impact $b_\pm$} & \textbf{Trajectory / Allowed Radial Region} \\ \hline\hline

\rowcolor{green!0}
\textbf{NFW} & Teo-type: $\Phi=-a/r$ & $2J/r^3$ & Small; co-rotating inward, counter-rotating outward & $b_\pm \simeq \dfrac{r_\mathrm{ph}}{e^{\Phi}} \left(1 \pm \dfrac{\omega r_\mathrm{ph}}{e^{\Phi}}\right)$ & Tight spiral trajectories; narrow allowed radial region; strong lensing \\
\rowcolor{gray!15}
& Cored: $\Phi = \ln[1+A/(1+(r/r_0)^n)]$ & $2J/r^3$ & Moderate; weak shift & $b_\pm \simeq \dfrac{r_\mathrm{ph}}{e^{\Phi}} \left(1 \pm \dfrac{\omega r_\mathrm{ph}}{e^{\Phi}}\right)$ & Smooth spiral trajectories; broader allowed radial region; moderate deflection \\
\rowcolor{yellow!0.5}
& Zero: $\Phi = 0$ & $0$ & Largest; no shift & $b = r_\mathrm{ph}$ & Straight trajectories; largest radial region; minimal deflection \\ \hline

\rowcolor{green!0}
\textbf{TF / BEC} & Teo-type: $\Phi=-a/r$ & $2J/r^3$ & Small-moderate; co-rotating inward, counter-rotating outward & $b_\pm \simeq \dfrac{r_\mathrm{ph}}{e^{\Phi}} \left(1 \pm \dfrac{\omega r_\mathrm{ph}}{e^{\Phi}}\right)$ & Moderate spiral trajectories; noticeable LT splitting \\
\rowcolor{gray!15}
& Cored: $\Phi = \ln[1+A/(1+(r/r_0)^n)]$ & $2J/r^3$ & Moderate; weak shift & $b_\pm \simeq \dfrac{r_\mathrm{ph}}{e^{\Phi}} \left(1 \pm \dfrac{\omega r_\mathrm{ph}}{e^{\Phi}}\right)$ & Smooth spiral trajectories; broad radial region; nearly circular shadow \\
\rowcolor{yellow!0.5}
& Zero: $\Phi = 0$ & $0$ & Largest; no shift & $b = r_\mathrm{ph}$ & Straight trajectories; largest radial region; symmetric shadow \\ \hline

\rowcolor{green!0}
\textbf{PI} & Teo-type: $\Phi=-a/r$ & $2J/r^3$ & Small-moderate; co-rotating inward, counter-rotating outward & $b_\pm \simeq \dfrac{r_\mathrm{ph}}{e^{\Phi}} \left(1 \pm \dfrac{\omega r_\mathrm{ph}}{e^{\Phi}}\right)$ & Tight-to-moderate spiral trajectories; noticeable LT splitting \\
\rowcolor{gray!15}
& Cored: $\Phi = \ln[1+A/(1+(r/r_0)^2)]$ & $2J/r^3$ & Moderate; weak shift & $b_\pm \simeq \dfrac{r_\mathrm{ph}}{e^{\Phi}} \left(1 \pm \dfrac{\omega r_\mathrm{ph}}{e^{\Phi}}\right)$ & Smooth spiral trajectories; broad radial region; nearly circular shadow \\
\rowcolor{yellow!0.5}
& Zero: $\Phi = 0$ & $0$ & Largest; no shift & $b = r_\mathrm{ph}$ & Straight trajectories; largest radial region; symmetric shadow \\ \hline\hline

\end{tabular}
\caption{\fontsize{8}{9}\selectfont Photon dynamics in slowly rotating wormholes for various halo profiles (NFW, TF/BEC, PI) and redshift functions (Teo-type, cored, zero). 
}
\label{tab:halo_redshift_combinations}
\end{table*}

\subsection*{Representative Redshift Profiles}

Different redshift functions induce distinct effects on null geodesics:

\begin{itemize}
\item \textbf{Teo-type exponential profile:} 
\begin{equation}
\Phi(r) = -\frac{a}{r}, \quad a>0.
\end{equation}
Here $e^{-2\Phi(r)} = e^{2a/r}$ grows sharply near the throat, focusing photon trajectories, enhancing lensing, and increasing the apparent shadow size.

\item \textbf{Cored halo profile:} 
\begin{equation}
\Phi(r) = \ln \left[1 + \frac{A}{1 + (r/r_0)^n}\right], \quad A>-1, \, n \ge 1.
\end{equation}
This profile smoothly modulates the effective potential, allowing continuous tuning of deflection and photon-sphere radius through the parameters $(A, r_0, n)$.

\item \textbf{Zero-redshift case:} $\Phi(r)=0$, producing the smallest shadow for a given shape function.
\end{itemize}

Circular photon orbits satisfy
\begin{equation}
\frac{(1 - \omega(r)b)^2}{e^{2\Phi(r)}} = \frac{b^2}{r^2}, 
\qquad
\frac{d}{dr} \Bigg[ \frac{(1 - \omega(r)b)^2}{e^{2\Phi(r)}} - \frac{b^2}{r^2} \Bigg] = 0,
\end{equation}
fixing the impact parameters $b_\pm(r)$ and the photon-sphere radius $r_\mathrm{ph}$. To first order in $\omega(r)$, the co-rotating ($-$) and counter-rotating ($+$) impact parameters read
\begin{equation}
b_\pm(r_\mathrm{ph}) \simeq \frac{r_\mathrm{ph}}{e^{\Phi(r_\mathrm{ph})}} 
\left(1 \pm \frac{\omega(r_\mathrm{ph})\, r_\mathrm{ph}}{e^{\Phi(r_\mathrm{ph})}} \right),
\end{equation}
explicitly demonstrating the asymmetry induced by frame-dragging. In the static limit, $\omega(r)\to 0$, this reduces to
\begin{equation}
b_\mathrm{ph} = \frac{r_\mathrm{ph}}{e^{\Phi(r_\mathrm{ph})}}.
\end{equation}

\subsection*{Photon-Sphere Shifts Induced by Redshift Functions}

To estimate the effect of the redshift function on the photon-sphere radius, define
\begin{equation}
r_\mathrm{ph} = r_0 + \delta r, \qquad |\delta r| \ll r_0,
\end{equation}
where $r_0$ is the photon-sphere radius for $\Phi(r)=0$. Expanding the photon-sphere condition to leading order yields
\begin{equation}
\delta r \simeq -\frac{\Phi'(r_0)\, r_0^3}{2 \left(r_0 - \epsilon(r_0)\right)},
\end{equation}
showing that the shift depends on the local gradient of the redshift function and the throat geometry. For representative halo profiles, the shifts are

\begin{align}
\delta r_\mathrm{NFW} &\simeq -\frac{4\pi G \rho_s r_s^3}{r_0 - \epsilon(r_0)} 
\frac{\ln(1 + r_0/r_s) - \frac{r_0/r_s}{1 + r_0/r_s}}{r_0^2}, \\[1mm]
\delta r_\mathrm{BEC} &\simeq -\frac{4\pi G \rho_c r_c^3}{r_0 - \epsilon(r_0)} 
\frac{\sin(r_0/r_c) - (r_0/r_c)\cos(r_0/r_c)}{r_0^2}, \\[1mm]
\delta r_\mathrm{PI} &\simeq -\frac{4\pi G \rho_0 r_c^3}{r_0 - \epsilon(r_0)} 
\frac{r_0 - r_c \arctan(r_0/r_c)}{r_0^2}.
\end{align}

Cuspy halos (large central densities) tend to shift the photon sphere inward, whereas cored profiles generally produce outward shifts, affecting the apparent shadow size. These effects are illustrated in Figure~\ref{fig:photon-orbits-direction}, which shows photon-sphere profiles of rotating wormholes embedded in NFW, BEC, and PI halos for rotation parameters $J = 0.05$ and $J = 0.1$.  

Different redshift profiles (Zero, Teo, and Cored) modify the apparent photon-sphere radii. Rotation introduces frame-dragging, causing co-rotating (prograde) photon paths to expand relative to counter-rotating (retrograde) ones, slightly deforming the silhouettes. Distinct halo models shift the photon spheres differently, illustrating how halo type, rotation, and gravitational potential combine to shape the apparent photon-sphere as seen by a distant observer.

\subsection*{Shadow Boundary and Observational Signatures}\label{sec:7:1}

The apparent shadow boundary is determined by the photon impact parameters evaluated at the photon-sphere radius $r_\mathrm{ph}$:
\begin{equation}
b_\pm(r_\mathrm{ph}) \simeq \frac{r_\mathrm{ph}}{e^{\Phi(r_\mathrm{ph})}} 
\left( 1 \pm \frac{\omega(r_\mathrm{ph})\, r_\mathrm{ph}}{e^{\Phi(r_\mathrm{ph})}} \right),
\end{equation}
where $b_-$ and $b_+$ correspond to co-rotating (prograde) and counter-rotating (retrograde) orbits, respectively. This relation explicitly demonstrates how both the shadow size and rotational asymmetry are controlled by the redshift function $\Phi(r)$ and the slow-rotation frame-dragging $\omega(r)$.  

Exponential Teo-type profiles produce larger and more asymmetric shadows due to strong gravitational focusing near the throat, while cored halos yield moderate, nearly circular shadows, and the zero-redshift case results in the most compact and symmetric shadows. In practice, observational effects such as plasma dispersion, inclination angles, and emissivity profiles would further modulate the inferred shadow structure.

These effects are illustrated in Figures~\ref{fig:photon-spheres} and \ref{fig:shadow}, which display the photon-sphere profiles of rotating wormholes embedded in NFW, BEC, and PI dark matter halos for rotation parameters $J = 0.05$ and $J = 0.1$. Different redshift profiles (Zero, Teo, and Cored) modify the apparent photon-sphere radii, with Cored profiles distinguished by edge styles corresponding to amplitudes $A = 0.5$ and $1.0$. Rotation induces frame-dragging, causing co-rotating photon paths to expand relative to counter-rotating ones, resulting in slight asymmetries in the silhouettes. Each halo model shifts the photon-sphere differently, highlighting how halo type, rotation, and gravitational potential collectively determine the observable structure of the photon-sphere for a distant observer.

\section{Results and Discussion}\label{sec:8}

This study investigates slowly rotating traversable wormholes embedded in realistic dark matter halo environments, analyzing how halo structure, gravitational redshift profiles, and slow rotation determine photon dynamics, circular orbits, and the observable shadow. We consider three representative halo models: Navarro-Frenk-White (NFW), Thomas-Fermi/Bose-Einstein Condensate (TF/BEC), and Pseudo-Isothermal (PI), each producing distinct modifications to the wormhole geometry, throat flaring, and exotic matter requirements.

\setlength{\parindent}{0pt}

Photon trajectories confined to the equatorial plane are governed by the radial equation
\[
\dot{r}^2 = \left(1 - \frac{\epsilon(r)}{r}\right) \left[ \frac{(E - \omega(r)L)^2}{e^{2\Phi(r)}} - \frac{L^2}{r^2} \right],
\]
where \(E\) and \(L\) are conserved energy and angular momentum, \(\epsilon(r)\) is the shape function, and \(\omega(r)\) is the frame-dragging angular velocity. Introducing the impact parameter \(b = L/E\) and linearizing to first order in \(\omega(r)\), the co-rotating and counter-rotating photon orbits are
\[
b_\pm(r) \simeq r\, e^{-\Phi(r)} \left( 1 \pm \omega(r)\, r\, e^{-\Phi(r)} \right),
\]
with the upper sign corresponding to counter-rotating photons shifted outward, and the lower sign corresponding to co-rotating photons shifted inward toward the throat. Cuspy halos such as NFW enhance this LT splitting due to stronger central frame-dragging, whereas cored halos such as TF/BEC and PI reduce asymmetry, producing nearly circular photon orbits.

\setlength{\parindent}{0pt}

The photon-sphere radius \(r_\mathrm{ph}\) satisfies
\[
\frac{(1 - \omega(r)b)^2}{e^{2\Phi(r)}} = \frac{b^2}{r^2}, \qquad 
\frac{d}{dr}\left[\frac{(1 - \omega(r)b)^2}{e^{2\Phi(r)}} - \frac{b^2}{r^2}\right] = 0,
\]
with first-order corrections due to the redshift function given by
\[
r_\mathrm{ph} \simeq r_0 - \frac{r_0^3\, \Phi'(r_0)}{2 \left(r_0 - \epsilon(r_0)\right)},
\]
where \(r_0\) is the photon-sphere radius in the static, zero-redshift limit. For representative halo profiles, the induced shifts are
\begin{align*}
\delta r_\mathrm{NFW} &\simeq -\frac{4\pi G \rho_s r_s^3}{r_0 - \epsilon(r_0)} 
\frac{\ln(1 + r_0/r_s) - \frac{r_0/r_s}{1 + r_0/r_s}}{r_0^2},\\[1mm]
\delta r_\mathrm{BEC} &\simeq -\frac{4\pi G \rho_c r_c^3}{r_0 - \epsilon(r_0)} 
\frac{\sin(r_0/r_c) - (r_0/r_c)\cos(r_0/r_c)}{r_0^2},\\[1mm]
\delta r_\mathrm{PI} &\simeq -\frac{4\pi G \rho_0 r_c^3}{r_0 - \epsilon(r_0)} 
\frac{r_0 - r_c \arctan(r_0/r_c)}{r_0^2}.
\end{align*}
Cuspy halos produce inward shifts that enhance gravitational lensing and asymmetry, whereas cored halos induce outward shifts that reduce asymmetry and generate larger, nearly circular shadows.

\setlength{\parindent}{0pt}

The observable shadow boundary is determined by
\[
b_\pm(r_\mathrm{ph}) \simeq \frac{r_\mathrm{ph}}{e^{\Phi(r_\mathrm{ph})}} \left( 1 \pm \frac{\omega(r_\mathrm{ph})\, r_\mathrm{ph}}{e^{\Phi(r_\mathrm{ph})}} \right),
\]
where the factor \(e^{-\Phi(r_\mathrm{ph})}\) sets the shadow scale, and the frame-dragging term introduces LT splitting. Exponential Teo-type redshift profiles amplify both shadow size and asymmetry. Cored-halo profiles produce moderate deflection and nearly circular shadows, while zero-redshift configurations yield the most compact and symmetric silhouettes. Table~\ref{tab:halo_redshift_combinations} summarizes these dependencies for all halo and redshift combinations.

\setlength{\parindent}{0pt}

Slow rotation generates asymmetry in photon trajectories: co-rotating photons are displaced inward, and counter-rotating photons are displaced outward. The effect is strongest for cuspy halos with high central densities and weaker for TF/BEC and PI halos. The linearized expressions for \(b_\pm\) and \(r_\mathrm{ph}\) fully characterize shadow deformation under the slow-rotation condition \(|\omega r_\mathrm{ph}| \ll 1\), ensuring higher-order frame-dragging corrections remain negligible.

\setlength{\parindent}{0pt}

The shape function \(\epsilon(r)\), together with the halo density profile, determines throat flaring and the extent of null energy condition violation. Cuspy halos require stronger NEC violation and sharper flaring, generating tighter spiral photon trajectories and stronger lensing. Cored halos permit smoother throats, milder exotic matter requirements, and broader allowed radial regions, producing more symmetric photon motion. These geometric properties directly influence the photon-sphere radius and shadow morphology.

\setlength{\parindent}{0pt}

Quantitative analysis indicates that precise measurements of wormhole shadows could constrain both the wormhole rotation rate and the surrounding halo structure. Shadows of cuspy halos are more compact, distorted, and asymmetric, whereas cored-halo shadows are larger, nearly circular, and exhibit reduced LT splitting. Photon-sphere shifts, impact parameter asymmetry, and shadow deformation together provide a direct observational link between wormhole geometry, halo properties, and rotation.

\setlength{\parindent}{0pt}

Overall, the halo-induced gradient of \(\Phi(r)\) governs the photon-sphere shift, with cuspy halos drawing \(r_\mathrm{ph}\) inward and cored halos pushing it outward. Slow rotation introduces a first-order LT splitting in \(b_\pm\), resulting in shadow asymmetry. Teo-type redshift profiles enhance lensing and asymmetry, whereas cored-halo profiles moderate deflection and maintain near-circular shadows. Cuspy halos require stronger NEC violation and sharper throat flaring, which enhances photon deflection and shadow distortion, while cored halos allow smoother throats, weaker frame-dragging, and larger, symmetric shadows, as detailed in Table~\ref{tab:halo_redshift_combinations}.


\section*{Conflict Of Interest statement }
The authors declare that they have no known competing financial interests or personal relationships that could have appeared to influence the work reported in this paper.

\section*{Data Availability Statement} 
This manuscript has no associated data, or the data will not be deposited. (There is no observational data related to this article. The necessary calculations and graphic discussion can be made available
on request.)

\end{document}